\documentclass[amsmath,amssymb,prl,superscriptaddress,showpacs,longbibliography,twocolumn]{revtex4-1}
\usepackage{graphicx}
\usepackage{dcolumn}
\usepackage[colorlinks=true,linkcolor=blue,citecolor=blue,urlcolor=blue]{hyperref}
\usepackage{stmaryrd}
\usepackage{esvect}
\usepackage[usenames,dvipsnames]{xcolor}
\usepackage{soul}

\newcommand{\MC}[1]{\mathcal{#1}}
\newcommand{\MR}[1]{\mathrm{#1}}
\newcommand{\VEC}[1]{\mathbf{#1}}

\usepackage{siunitx}
\DeclareSIUnit\mub{\mu_\text{B}}
\usepackage{mathtools}

\DeclarePairedDelimiterX\braket[1]{\langle}{\rangle}{#1}

\begin{document}

\title{Proper and improper chiral magnetic interactions}

\author{Manuel dos Santos Dias}\email{m.dos.santos.dias@fz-juelich.de}
\affiliation{Peter Gr\"{u}nberg Institut and Institute for Advanced Simulation, Forschungszentrum J\"{u}lich \& JARA, 52425 J\"{u}lich, Germany}
\author{Sascha Brinker}\email{s.brinker@fz-juelich.de}
\affiliation{Peter Gr\"{u}nberg Institut and Institute for Advanced Simulation, Forschungszentrum J\"{u}lich \& JARA, 52425 J\"{u}lich, Germany}
\author{Andr\'as L\'aszl\'offy}
\affiliation{Wigner Research Centre for Physics, P.O. Box 49, H-1525 Budapest, Hungary}
\affiliation{Department of Theoretical Physics, Budapest University of Technology and Economics, Budafoki \'ut 8, H-1111 Budapest, Hungary}
\author{Bendeg\'uz Ny\'ari}
\affiliation{Department of Theoretical Physics, Budapest University of Technology and Economics, Budafoki \'ut 8, H-1111 Budapest, Hungary}
\author{Stefan Bl\"ugel}
\affiliation{Peter Gr\"{u}nberg Institut and Institute for Advanced Simulation, Forschungszentrum J\"{u}lich \& JARA, 52425 J\"{u}lich, Germany}
\author{L\'aszl\'o Szunyogh}
\affiliation{Department of Theoretical Physics, Budapest University of Technology and Economics, Budafoki \'ut 8, H-1111 Budapest, Hungary}
\affiliation{MTA-BME Condensed Matter Research Group, Budapest University of Technology and Economics, Budafoki \'ut 8, H-1111 Budapest, Hungary}
\author{Samir Lounis}\email{s.lounis@fz-juelich.de}
\affiliation{Peter Gr\"{u}nberg Institut and Institute for Advanced Simulation, Forschungszentrum J\"{u}lich \& JARA, 52425 J\"{u}lich, Germany}
\affiliation{Faculty of Physics, University of Duisburg-Essen and CENIDE, 47053 Duisburg, Germany}

\date{\today}

\begin{abstract}
Atomistic spin models are of great value for predicting and understanding the magnetic properties of real materials, and extensions of the existing models open routes to new physics and potential applications.
The Dzyaloshinskii-Moriya interaction is the prototype for chiral magnetic interactions, and several recent works have uncovered or proposed various types of generalized chiral interactions.
However, in some cases the proposed interactions or their interpretation do not comply with basic principles such as being independent of the magnetic configuration from which they are evaluated, or even obeying time-reversal invariance.
In this brief contribution, we present a simple explanation for the origin of these puzzling findings, and point out how to resolve them.
\end{abstract}

\maketitle

\emph{Introduction.}
The magnetic interaction between two spin moments is partitioned in isotropic, anisotropic, scalar and vector chiral interactions. The latter is commonly referred to as the chiral magnetic interaction and  was developed by Dzyaloshinskii and Moriya~\cite{Dzyaloshinskii1957,Moriya1960}.
What is commonly understood by Dzyaloshinskii-Moriya interaction (DMI) is an antisymmetric exchange interaction that can be written for two spin moments as $\VEC{D}_{12}\cdot\left(\VEC{S}_1 \times \VEC{S}_2\right)$.
The interaction vector $\VEC{D}_{12}$ obeys the symmetry rules enumerated by Moriya~\cite{Moriya1960}.
Different microscopic mechanism have been identified that lead to the DMI~\cite{Moriya1960,Smith1976,Fert1980,Imamura2004,Kikuchi2016}, all having in common the need for the relativistic spin-orbit interaction and an inversion asymmetric environment.
The DMI underpins many interesting magnetic systems, such as weak ferromagnetism in antiferromagnets~\cite{Dzyaloshinskii1957}, spin spiral ground states~\cite{Bode2007}, chiral magnetic domain walls~\cite{Emori2013}, and magnetic skyrmions~\cite{Nagaosa2013,Fert2017}.

With the advent of realistic electronic structure calculations, it became possible to compute the magnitude and other properties of the magnetic exchange interactions for a given material.
A real-space approach to the pairwise interactions (the infinitesimal rotation method) was introduced in Ref.~\cite{Liechtenstein1987} and later generalized for the calculation of the DMI and other anisotropic interactions~\cite{Udvardi2003,Ebert2009}.
The central idea is to establish a mapping between the electronic structure calculations and a suitably-defined atomistic spin model, but soon it was realized that this mapping is not trivial and can lead to a dependence of the computed interactions on the reference magnetic configuration~\cite{Lounis2010a,Szilva2013,Szilva2017}.
A large part of these dependencies can be accounted for by expanding the reference atomistic spin model to include not only Heisenberg pairwise interactions but also biquadratic interactions~\cite{Mryasov1996,Lounis2010a,Szilva2013}, leading to generalized infinitesimal rotation approaches to multi-spin interactions~\cite{Mankovsky2020,Lounis2020}.

A different approach consists in computing the energies of different magnetic configurations and parametrizing a spin model that can reproduce these energies.
The general framework is called the spin cluster expansion and is easily applied to systematically map the complete set of interactions for a finite number of magnetic moments~\cite{Drautz2004,Drautz2005,Antal2008,Singer2011}.
We have recently implemented the self-consistent spin cluster expansion within a constrained density functional theory framework~\cite{Brinker2019,Brinker2020a}, leading to the identification of those four-spin interactions that are counterparts of the DMI: the chiral biquadratic interaction~\cite{Brinker2019} in magnetic dimers, and its multi-site counterparts in trimers and tetramers~\cite{Brinker2020a}, with some of these interactions not being constrained by Moriya's rules.
The significance of the chiral multi-spin interactions has also been recognized in~\cite{Laszloffy2019,Grytsiuk2020,Mankovsky2020}.

Given the large interest in chiral magnetic interactions, we note that a few recent works have advanced unwarranted interpretations of otherwise sound first-principles calculations, such as chiral three-spin interactions that are incompatible with time-reversal symmetry~\cite{Mankovsky2020}, or a very large DMI~\cite{Cardias2020,Cardias2020a} that depends strongly on the magnetic configuration and does not rely on the spin-orbit interaction.
The latter contradicts the understanding of the DMI as established by the magnetism community over the past 60 years, without offering a compelling theoretical justification for its revision.
The purpose of this paper is to present a simple explanation for these proposed chiral interactions that we term `improper', complying with basic symmetry requirements on the atomistic spin model.

\emph{Proper and improper magnetic interactions.}
Before we discuss our results, we define the terminology that we use in this work.
We term as `proper' those magnetic interactions which are parametrized by interaction coefficients which are independent of the magnetic state of the system and are invariant under time-reversal symmetry (so they consist of an even number of spins), and if these properties are not satisfied we use the term `improper'.
It is always possible to give a complete parametrization of the magnetic energy using only proper magnetic interactions~\cite{Drautz2004,Drautz2005,Antal2008,Singer2011,Brinker2019,Brinker2020a}.
For example, our analysis of a generic electronic model in Appendix B of Ref.~\cite{Brinker2019} shows that only terms with an even number of magnetic moments appear in an infinite series expansion of the electronic grand potential (so no three-spin terms as advocated in Ref.~\cite{Mankovsky2020}), and that terms containing only one explicit cross-product between them require the spin-orbit interaction (unlike the effective DMI of Refs.~\cite{Cardias2020,Cardias2020a}).

\begin{figure*}[!t]
    \centering
    \includegraphics[width=0.8\textwidth]{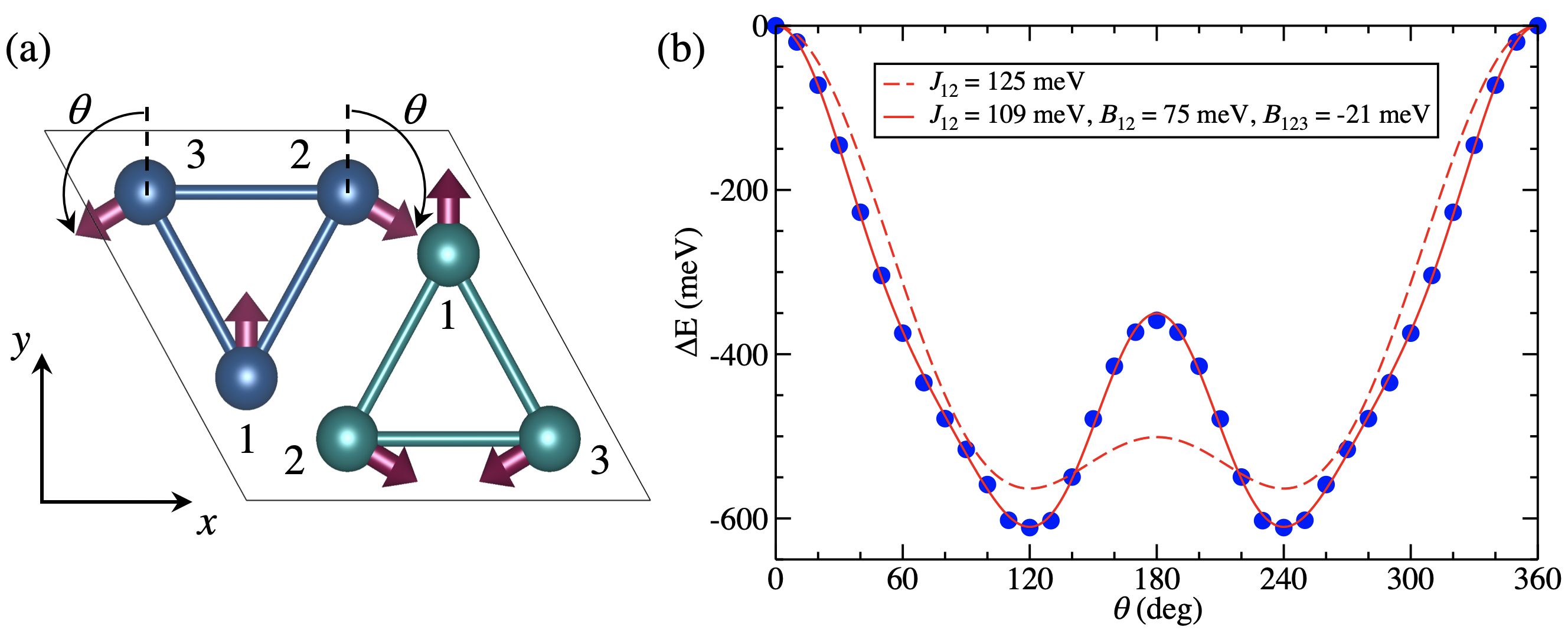}
    \caption{\label{fig:uppsala}
    Dependence of the total energy of Mn$_3$Sn on the magnetic configuration.
    (a) Top view of the unit cell of Mn$_3$Sn with noncollinear magnetic structure viewed along the $z$-axis.
    The spheres indicate the Mn atoms on the $z = 1/4$ plane (blue) and on the $z=3/4$ plane (green).
    Sn is not shown.
    (b) Total energy for the magnetic configurations shown in (a)
    fitted to the spin model of Eq.~\eqref{eq:trimer_model} including only isotropic two-sublattice ($J_{12}$) or also biquadratic and three-sublattice interactions ($B_{12}$ and $B_{123}$, respectively).
    }
\end{figure*}

\emph{Improper DMI in Mn$_3$Sn.}
We start by reinterpreting the first-principles calculations performed for Mn$_3$Sn in Ref.~\cite{Cardias2020}.
This compound belongs to the space group $P6_3/mmc$ (194).
The crystallographic unit cell shown in Fig.~\ref{fig:uppsala}(a) consists of six Mn atoms that are partitioned into three ferromagnetic sublattices denoted by the indices $i,j\in\{1,2,3\}$.
Each sublattice contains two atoms in the unit cell, one at $z=1/4$ and one at $z=3/4$.
Ref.~\cite{Cardias2020} computed the total energies for the family of magnetic configurations obtained when the magnetization for sublattices 2 and 3 are rotated by an angle $\pm \theta$ relative to that in sublattice 1.
Hence, $\theta = \SI{0}{\degree}$ denotes the ferromagnetic configuration, $\theta = \SI{120}{\degree}$ denotes the triangular N\'eel configuration shown in Fig.~\ref{fig:uppsala}(a), and $\theta = \SI{180}{\degree}$ is a ferrimagnetic up-down-down state.

We repeated the total energy calculations of Ref.~\cite{Cardias2020} employing  the all-electron Korringa-Kohn-Rostoker Green function method in full potential~\cite{Papanikolaou:2002} with spin-orbit coupling added to the scalar relativistic approximation~\cite{Bauer2014}. We performed self-consistent calculations without constraints by treating  exchange and correlation effects in the local spin density approximation~\cite{Vosko1980}.
We adopt the experimental lattice geometry following Ref.~\cite{Tomiyoshi1982}.
The scattering wave functions are expanded up to an angular momentum cutoff of $\ell_\text{max}=3$ and a k-mesh of $24 \times 24\times 30$ is used.
Our results shown in Fig.~\ref{fig:uppsala}(b) are consistent with Ref.~\cite{Cardias2020}, exhibiting a pronounced fourth-order type energy behavior with respect to the rotation of the angle $\theta$.
Without spin-orbit interaction the antiferromagnetic triangular N\'eel states at $\theta=120^\circ$ and $\theta=240^\circ$ are degenerate.
Including the spin-orbit interaction the well-known DMI emerges and the magnetic structure with $\Gamma_5$ symmetry corresponding to $\theta=240^\circ$ is lower in energy by \SI{8}{\milli\electronvolt} than the one with $\Gamma_3$ symmetry denoted by $\theta=120^\circ$, in agreement with the discussion of Ref.~\cite{Nyari2019}.

We depart from the analysis in Ref.~\cite{Cardias2020} by fitting the dependence of our total energies on the magnetic configuration of the Mn sublattices $i$ according to the extended spin model that we presented in Refs.~\cite{Hoffmann2020,Brinker2020a},
($|\VEC{S}_i| = 1$):
\begin{align}
    \label{eq:trimer_model}
    \MC{E} &= J_{12} \left(\VEC{S}_1\cdot\VEC{S}_2 + \VEC{S}_2\cdot\VEC{S}_3 + \VEC{S}_3\cdot\VEC{S}_1\right) \nonumber\\
    &+ B_{12} \left((\VEC{S}_1\cdot\VEC{S}_2)^2 + (\VEC{S}_2\cdot\VEC{S}_3)^2 + (\VEC{S}_3\cdot\VEC{S}_1)^2\right) \nonumber\\
    &+ B_{123} \left((\VEC{S}_1\cdot\VEC{S}_2)(\VEC{S}_2\cdot\VEC{S}_3) + (\VEC{S}_2\cdot\VEC{S}_3)(\VEC{S}_3\cdot\VEC{S}_1)\right. \nonumber\\
    &\left.\hspace{3.5em}+ (\VEC{S}_3\cdot\VEC{S}_1)(\VEC{S}_1\cdot\VEC{S}_2)\right) \;.
\end{align}
Here we define the bilinear isotropic interaction between two sublattices with strength $J_{12}$, and two types of four-spin isotropic interactions between the sublattices: biquadratic with strength $B_{12}$ and three-sublattice with strength $B_{123}$.
Fig.~\ref{fig:uppsala}(b) shows that the fit using only the bilinear two-sublattice interactions is quite poor, as claimed in Ref.~\cite{Cardias2020}, while inclusion of the four-spin two- and three-sublattice interactions fits the full angular dependence almost perfectly.

To see how the interpretation in terms of an improper Dzyaloshinskii-Moriya sublattice interaction can come about, we follow the strategy of Ref.~\cite{Cardias2020} and derive the first-order change in the magnetic energy due to a small deviation of one of the spin directions for the spin model of Eq.~\eqref{eq:trimer_model}.
We replace $\VEC{S}_1 \rightarrow \VEC{S}_1 + \delta\VEC{S}_1$ and find how the energy given by Eq.~\ref{eq:trimer_model} changes due to this perturbation, with the result $\delta\MC{E}_1 = \delta\MC{E}^\MR{J}_1 + \delta\MC{E}^\MR{B}_1$.
The two contributions are
\begin{equation}
    \delta\MC{E}^\MR{J}_1 = J_{12} \left(\VEC{S}_2 + \VEC{S}_3\right) \cdot \delta\VEC{S}_1
\end{equation}
\begin{equation}
    \delta\MC{E}^\MR{B}_1 = 2B_{12}\,\big( (\VEC{S}_1\cdot\VEC{S}_2)\,\VEC{S}_2 + (\VEC{S}_3\cdot\VEC{S}_1)\,\VEC{S}_3\big) \cdot \delta\VEC{S}_1 \;.
\end{equation}
The other four-spin interaction is discussed later.
We have nothing to remark about $\delta\MC{E}^\MR{J}_1$, but $\delta\MC{E}^\MR{B}_1$ can be expressed in an alternative form using the vector identity
\begin{equation}\label{eq:doublecross}
    (\VEC{A}\times\VEC{B})\cdot(\VEC{C}\times\VEC{D}) = (\VEC{A}\cdot\VEC{C})\,(\VEC{B}\cdot\VEC{D}) - (\VEC{A}\cdot\VEC{D})\,(\VEC{B}\cdot\VEC{C}) \;.
\end{equation}
For small deviations $\VEC{S}_1\cdot\delta\VEC{S}_1 = 0$, and we find
\begin{align}
    \delta\MC{E}^\MR{B}_1 &= 2B_{12}\,(\VEC{S}_1\times\VEC{S}_2) \cdot (\VEC{S}_2\times\delta\VEC{S}_1) \nonumber\\
    &+ 2B_{12}\,(\VEC{S}_1\times\VEC{S}_3) \cdot (\VEC{S}_3\times\delta\VEC{S}_1) \nonumber\\
    &= \VEC{D}_{21} \cdot (\VEC{S}_2\times\delta\VEC{S}_1)
    + \VEC{D}_{31} \cdot (\VEC{S}_3\times\delta\VEC{S}_1) \;.
\end{align}
These effective DM vectors are artificial, as they were obtained by rearranging the contribution from the isotropic four-spin sublattice interactions.
Expanding their definition, e.g.
\begin{equation}
    \VEC{D}_{21} = 2B_{12}\,(\VEC{S}_1\times\VEC{S}_2) = 2B_{12} \sin\theta\,\VEC{z} = D_{21}(\theta)\,\VEC{z} \;.
\end{equation}
It follows that these effective DM vectors strongly depend on the magnetic configuration chosen for their calculation.
They vanish for collinear magnetic configurations ($\theta = \SI{0}{\degree},\SI{180}{\degree}$), have a maximum magnitude of $2B_{12} \approx \SI{150}{\milli\electronvolt}$, and do not arise from the spin-orbit interaction.
Lastly, they are achiral, as chirality reversal is achieved by $\theta \rightarrow -\theta$ which leads to $D_{21}(-\theta) = -D_{21}(\theta)$, and so both chiralities have the same energy.
All these properties match those reported in Ref.~\cite{Cardias2020} including the magnitude of the interaction (see Fig.~3 within that reference), which points to isotropic four-spin sublattice interactions as the origin of the effective DMI proposed in that work.

\emph{Improper DMI in magnetic trimers.}
The authors of Ref.~\cite{Cardias2020} have recently published a related work~\cite{Cardias2020a} which applies the same reasoning to magnetic Cr and Mn trimers on the Ag(111) and Au(111) surfaces.
One methodological difference is that the improper DMI was computed by considering changes in two spins simultaneously, e.g.\ $\delta\VEC{S}_1$ and $\delta\VEC{S}_2$.
Our spin model from Eq.~\eqref{eq:trimer_model} in combination with the vector identity of Eq.~\eqref{eq:doublecross} leads to the second-order variations of the Heisenberg interaction
\begin{equation}
    \delta\MC{E}^\MR{J}_{12} = J_{12}\, \delta\VEC{S}_1 \cdot \delta\VEC{S}_2 \;,
\end{equation}
and of the biquadratic and three-site interactions
\begin{align}\label{eq:dmi2}
    \delta\MC{E}^\MR{B}_{12} &= 2B_{12}\,(\VEC{S}_1\cdot\VEC{S}_2)\,(\delta\VEC{S}_1 \cdot \delta\VEC{S}_2) \nonumber\\
    &+ B_{123}\,(\VEC{S}_1+\VEC{S}_2)\cdot\VEC{S}_3\,(\delta\VEC{S}_1 \cdot \delta\VEC{S}_2)  \nonumber\\
    &+ 2B_{12}\,(\VEC{S}_1\times\VEC{S}_2)\cdot(\delta\VEC{S}_2 \times \delta\VEC{S}_1) \nonumber\\
    &+ B_{123}\,(\VEC{S}_1\times\VEC{S}_3 + \VEC{S}_3 \times \VEC{S}_2) \cdot (\delta\VEC{S}_2 \times \delta\VEC{S}_1) \nonumber\\
    &+ B_{123}\,(\delta\VEC{S}_1\cdot\VEC{S}_3)\,(\VEC{S}_3 \cdot \delta\VEC{S}_2)
    \;.
\end{align}
The first and second lines are contributions to an effective Heisenberg-like interaction, the third and fourth lines to an improper DMI-like interaction, and the last line to an artificial Ising-like interaction.
This shows that the effective interactions obtained by varying two spin orientations can also generate contributions which are hard to interpret without comparing to an appropriate spin model, and can depend not only on the spin alignment for the pair of interest (spins 1 and 2) but also on how they align with the rest of the system (spin 3).
We propose that these interactions explain the very large improper DMI reported in Table 2 of Ref.~\cite{Cardias2020a} in contrast with the proper one driven by the spin-orbit interaction.

Unfortunately, a direct quantitative comparison between Ref.~\cite{Cardias2020a} and our published results for Cr and Mn trimers on Au(111) in \cite{Antal2008} and \cite{Brinker2020a} is not possible due to large computational differences.
For the benefit of the reader, we note that the magnetic interactions were evaluated in Ref.~\cite{Antal2008} by fitting the energies of a large number of magnetic configurations in combination with the magnetic force theorem, while in \cite{Brinker2020a} they were fitted to the self-consistent constraining fields stabilizing a selected number of symmetry-inequivalent magnetic configurations.
For a quick comparison between the magnitudes of the proper and improper DMI using our own results, we select the Cr trimer on Au(111) and adopt the triangular N\'eel state.
In the notation of~\cite{Brinker2020a}, the $z$-component of the improper DMI for this magnetic configuration is $D_{21}^z = \sqrt{3}\left(B_{12} - B_{123}\right)$, which amounts to $\SI{-20}{\milli\electronvolt}$~\cite{Antal2008} or $\SI{-23}{\milli\electronvolt}$~\cite{Brinker2020a}.
The improper DMI is much larger than the proper DMI, $\SI{0.97}{\milli\electronvolt}$~\cite{Antal2008} or $\SI{1.7}{\milli\electronvolt}$~\cite{Brinker2020a}, but still more than six times smaller than the value of $\SI{134}{\milli\electronvolt}$ reported by Ref.~\cite{Cardias2020}.
We find it hard to explain such a large discrepancy solely with computational differences between our two works and Ref.~\cite{Cardias2020}.
As Refs.~\cite{Antal2008} and \cite{Brinker2020a} use a systematic mapping of the energy as a function of the magnetic configuration, such a large missing contribution to the energy would have been noticed.

\begin{figure}[!t]
    \centering
    \includegraphics[width=\columnwidth]{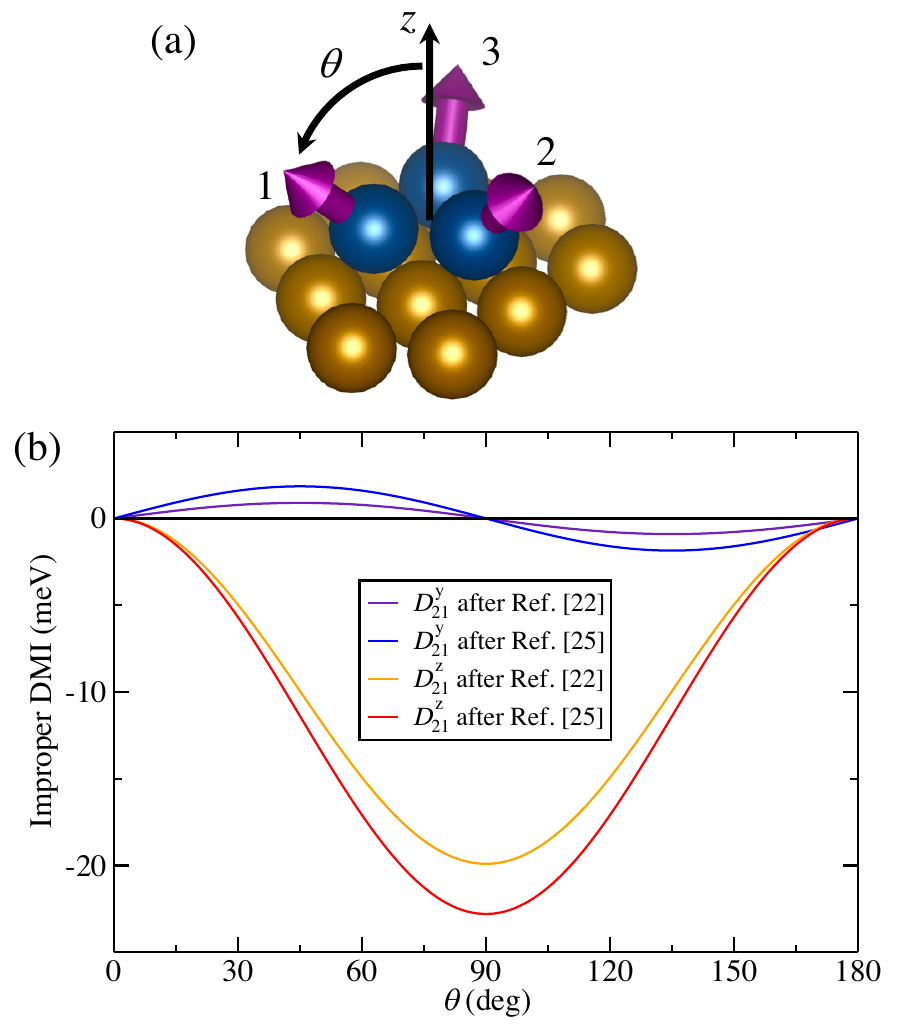}
    \caption{\label{fig:trimer}
    Dependence of the improper DMI on the magnetic configuration for a Cr trimer on Au(111).
    (a) Noncollinear magnetic structure.
    The blue spheres represent Cr atoms and the golden spheres Au atoms.
    (b) Dependence of the improper DMI between atoms 1 and 2 due to isotropic four-spin interactions on the magnetic configuration, according to Eq.~\eqref{eq:dmi_trimer} and the parameters $B_{12} = \SI{-4.42}{\milli\electronvolt}$ and $B_{123} = \SI{7.06}{\milli\electronvolt}$ from Ref.~\cite{Antal2008} and $B_{12} = \SI{-5.10}{\milli\electronvolt}$ and $B_{123} = \SI{8.06}{\milli\electronvolt}$ from Ref.~\cite{Brinker2020a}.
    }
\end{figure}

We now look at the dependence of the artificial DMI on a different set of magnetic configurations~\cite{Brinker2020a,Cardias2020a}, as illustrated in Fig.~\ref{fig:trimer}(a):
\begin{subequations}\label{eq:spin_structure_trimer}
\begin{align}
    \VEC{S}_1 &= \left(\frac{\sqrt{3}}{2}\sin\theta,\frac{1}{2}\sin\theta,\cos\theta\right) \;, \\
    \VEC{S}_2 &= \left(-\frac{\sqrt{3}}{2}\sin\theta,\frac{1}{2}\sin\theta,\cos\theta\right) \;, \\
    \VEC{S}_3 &= \left(0,-\sin\theta,\cos\theta\right) \;.
\end{align}
\end{subequations}
These keep the threefold rotation symmetry about the $z$-axis.
Here $\theta = \SI{0}{\degree}$ is a ferromagnetic arrangement with the spin moments pointing along $+z$, $\theta = \SI{90}{\degree}$ is the triangular N\'eel configuration, and $\theta = \SI{180}{\degree}$ is a ferromagnetic arrangement with the spin moments pointing along $-z$.
The improper DMI for a representative pair follows from Eq.~\eqref{eq:dmi2} and the chosen magnetic configuration:
\begin{subequations}\label{eq:dmi_trimer}
\begin{align}
    D_{21}^y(\theta) &= -\sqrt{3} \left(B_{12} + \frac{1}{2} B_{123}\right) \sin2\theta \;,\\
    D_{21}^z(\theta) &= \sqrt{3} \left(B_{12} - B_{123}\right) \sin^2\theta \;.
\end{align}
\end{subequations}
This shows that the same isotropic four-spin interactions can lead to different components of the improper DMI, depending on the reference magnetic configuration.
The angular dependence of the improper DMI is shown in Fig.~\ref{fig:trimer}(b), using the data from Refs.~\cite{Antal2008,Brinker2020a}.

\emph{Improper three-spin chiral interactions.}
We have argued so far that isotropic four-spin interactions can be misleadingly interpreted as improper chiral interactions.
However, we have recently identified proper chiral four-spin interactions in several independent works~\cite{Laszloffy2019,Grytsiuk2020,Brinker2020a}, so we now explain how they fit the story.
The concept of effective two-spin interactions is useful to characterize the local energy landscape around a reference magnetic configuration.
In this context, the isotropic four-spin interactions can be shown to renormalize the isotropic Heisenberg interaction while the chiral four-spin interactions renormalize the conventional DMI~\cite{Brinker2020a}.

These chiral four-spin interactions can also explain the improper chiral three-spin interactions introduced in Ref.~\cite{Mankovsky2020}.
There the infinitesimal rotation method was extended to a multi-spin multi-site framework, which among other things was used to identify the energy change from introducing spin deviations on three distinct sites.
We stress that we do not doubt the numerical results of Ref.~\cite{Mankovsky2020}, but point out that their proposed interpretation with a chiral three-spin three-site interaction built out of the scalar spin chirality of the perturbed moments, $\delta\VEC{S}_1\cdot\left(\delta\VEC{S}_2\times\delta\VEC{S}_3\right)$, is manifestly inconsistent with time-reversal symmetry.
They remedied this inconsistency by defining a magnetic-configuration-dependent interaction strength, in a way similar to the previously-discussed improper DMI.
The origin of these improper three-spin interactions can be found by starting from the form of the chiral four-spin interactions given in Ref.~\cite{Grytsiuk2020},
\begin{equation}
    \MC{E}^\MR{C} = \left(\VEC{C}_{123}\cdot\VEC{S}_1\right) \VEC{S}_1\cdot\left(\VEC{S}_2\times\VEC{S}_3\right) + \ldots
\end{equation}
The dots denote other terms obtained by permutations of $\{1,2,3\}$.
Note that this form of a chiral three-site interaction incorporates some symmetries of the underlying lattice and cannot describe the general case as discussed in Ref.~\cite{Brinker2020a}.
If we introduce small deviations of the three spin orientations simultaneously, we arrive at 
\begin{align}
    \Delta\MC{E}^\MR{C}_{123} &= \left(\VEC{C}_{123}\cdot\VEC{S}_1\right) \delta\VEC{S}_1\cdot\left(\delta\VEC{S}_2\times\delta\VEC{S}_3\right) + \ldots \\
    &= J_{123}^{(1)}\,\delta\VEC{S}_1\cdot\left(\delta\VEC{S}_2\times\delta\VEC{S}_3\right) + \ldots
\end{align}
where $J_{123}^{(1)}$ is a contribution to the chiral three-spin interaction coefficient in the notation of Ref.~\cite{Mankovsky2020}. 
It depends on the proper chiral four-spin interaction vector $\VEC{C}_{123}$~\cite{Laszloffy2019,Grytsiuk2020,Brinker2020a} and how the reference magnetic configuration is aligned with it.
We see no reason to introduce and use such improper chiral three-spin interactions when the proper chiral four-spin ones have already been uncovered and described.
Surprisingly, the authors of Ref.~\cite{Mankovsky2020} also acknowledge this in practice, as stated in the text following Eq.\ 29 within that work.

\emph{Conclusions.}
We showed how isotropic four-spin interactions can be misinterpreted as improper DMI-like interactions that have very pathological properties, chiefly being strongly dependent on the reference magnetic configuration used in the calculation.
The `natural' choice of a collinear ferromagnetic state as the reference state avoids most of these problems due to all magnetic moments being collinear.
Great care must be exercised in interpreting the results of the infinitesimal rotation approach if the reference state is not collinear, and we propose that for this to be successful the appropriate spin model must be defined beforehand in accordance with all the symmetries of the material (see Ref.~\cite{Brinker2020a} for a detailed discussion).
Lastly, we remark that improper chiral three-spin three-site interactions can likewise be obtained from varying the orientations of three magnetic moments and misinterpreting the result~\cite{Mankovsky2020}.
These arise from proper chiral four-spin interactions~\cite{Laszloffy2019,Grytsiuk2020,Brinker2020a}, which interestingly were also identified in Ref.~\cite{Mankovsky2020}, and acquire a dependence on the magnetic configuration through the orientation of the undisturbed fourth spin.

\emph{Acknowledgements.}
We gratefully acknowledge financial support from the DARPA TEE program through grant MIPR (\# HR0011831554) from DOI, from the European Research Council (ERC) under the European Union's Horizon 2020 research and innovation program (Grant No.\ 856538, project ``3D MAGiC'' and ERC-consolidator grant 681405 — DYNASORE),  from Deutsche For\-schungs\-gemeinschaft (DFG) through SPP 2137 ``Skyrmionics'' (Project BL 444/16), Priority Programme SPP 2244 ``2D Materials - Physics of van der Waals Heterostructures'' (project LO 1659/7-1), and the Collaborative Research Centers SFB 1238 (Project C01).
B.N., A.L. and L.S. acknowledge the support provided by the Ministry of Innovation and the National Research, Development, and Innovation (NRDI)  Office under Project No. K131938, by the NRDI Fund (TKP2020 IES, Grant No. BME-IE-NAT) and within the Quantum Information National Laboratory of Hungary.
The authors gratefully acknowledge the computing time granted through JARA-HPC on the supercomputer JURECA at the Forschungszentrum J\"ulich~\cite{jureca}.

\bibliography{references.bib}

\begin{thebibliography}{36}%
\makeatletter
\providecommand \@ifxundefined [1]{%
 \@ifx{#1\undefined}
}%
\providecommand \@ifnum [1]{%
 \ifnum #1\expandafter \@firstoftwo
 \else \expandafter \@secondoftwo
 \fi
}%
\providecommand \@ifx [1]{%
 \ifx #1\expandafter \@firstoftwo
 \else \expandafter \@secondoftwo
 \fi
}%
\providecommand \natexlab [1]{#1}%
\providecommand \enquote  [1]{``#1''}%
\providecommand \bibnamefont  [1]{#1}%
\providecommand \bibfnamefont [1]{#1}%
\providecommand \citenamefont [1]{#1}%
\providecommand \href@noop [0]{\@secondoftwo}%
\providecommand \href [0]{\begingroup \@sanitize@url \@href}%
\providecommand \@href[1]{\@@startlink{#1}\@@href}%
\providecommand \@@href[1]{\endgroup#1\@@endlink}%
\providecommand \@sanitize@url [0]{\catcode `\\12\catcode `\$12\catcode
  `\&12\catcode `\#12\catcode `\^12\catcode `\_12\catcode `\%12\relax}%
\providecommand \@@startlink[1]{}%
\providecommand \@@endlink[0]{}%
\providecommand \url  [0]{\begingroup\@sanitize@url \@url }%
\providecommand \@url [1]{\endgroup\@href {#1}{\urlprefix }}%
\providecommand \urlprefix  [0]{URL }%
\providecommand \Eprint [0]{\href }%
\providecommand \doibase [0]{http://dx.doi.org/}%
\providecommand \selectlanguage [0]{\@gobble}%
\providecommand \bibinfo  [0]{\@secondoftwo}%
\providecommand \bibfield  [0]{\@secondoftwo}%
\providecommand \translation [1]{[#1]}%
\providecommand \BibitemOpen [0]{}%
\providecommand \bibitemStop [0]{}%
\providecommand \bibitemNoStop [0]{.\EOS\space}%
\providecommand \EOS [0]{\spacefactor3000\relax}%
\providecommand \BibitemShut  [1]{\csname bibitem#1\endcsname}%
\let\auto@bib@innerbib\@empty
\bibitem [{\citenamefont {Dzyaloshinskii}(1957)}]{Dzyaloshinskii1957}%
  \BibitemOpen
  \bibfield  {author} {\bibinfo {author} {\bibfnamefont {I.~E.}\ \bibnamefont
  {Dzyaloshinskii}},\ }\bibfield  {title} {\enquote {\bibinfo {title}
  {Thermodynamic theory of "weak" ferromagnetism in antiferromagnetic
  substances},}\ }\href@noop {} {\bibfield  {journal} {\bibinfo  {journal}
  {Sov. Phys. JETP}\ }\textbf {\bibinfo {volume} {5}},\ \bibinfo {pages} {1259
  -- 1272} (\bibinfo {year} {1957})}\BibitemShut {NoStop}%
\bibitem [{\citenamefont {Moriya}(1960)}]{Moriya1960}%
  \BibitemOpen
  \bibfield  {author} {\bibinfo {author} {\bibfnamefont {Toru}\ \bibnamefont
  {Moriya}},\ }\bibfield  {title} {\enquote {\bibinfo {title} {Anisotropic
  superexchange interaction and weak ferromagnetism},}\ }\href {\doibase
  10.1103/PhysRev.120.91} {\bibfield  {journal} {\bibinfo  {journal} {Phys.
  Rev.}\ }\textbf {\bibinfo {volume} {120}},\ \bibinfo {pages} {91--98}
  (\bibinfo {year} {1960})}\BibitemShut {NoStop}%
\bibitem [{\citenamefont {Smith}(1976)}]{Smith1976}%
  \BibitemOpen
  \bibfield  {author} {\bibinfo {author} {\bibfnamefont {D.A.}\ \bibnamefont
  {Smith}},\ }\bibfield  {title} {\enquote {\bibinfo {title} {New mechanisms
  for magnetic anisotropy in localised s-state moment materials},}\ }\href
  {\doibase 10.1016/0304-8853(76)90069-X} {\bibfield  {journal} {\bibinfo
  {journal} {J. Magn. Magn. Mater.}\ }\textbf {\bibinfo {volume} {1}},\
  \bibinfo {pages} {214 -- 225} (\bibinfo {year} {1976})}\BibitemShut {NoStop}%
\bibitem [{\citenamefont {Fert}\ and\ \citenamefont {Levy}(1980)}]{Fert1980}%
  \BibitemOpen
  \bibfield  {author} {\bibinfo {author} {\bibfnamefont {A.}~\bibnamefont
  {Fert}}\ and\ \bibinfo {author} {\bibfnamefont {Peter~M.}\ \bibnamefont
  {Levy}},\ }\bibfield  {title} {\enquote {\bibinfo {title} {Role of
  anisotropic exchange interactions in determining the properties of
  spin-glasses},}\ }\href {\doibase 10.1103/PhysRevLett.44.1538} {\bibfield
  {journal} {\bibinfo  {journal} {Phys. Rev. Lett.}\ }\textbf {\bibinfo
  {volume} {44}},\ \bibinfo {pages} {1538--1541} (\bibinfo {year}
  {1980})}\BibitemShut {NoStop}%
\bibitem [{\citenamefont {Imamura}\ \emph {et~al.}(2004)\citenamefont
  {Imamura}, \citenamefont {Bruno},\ and\ \citenamefont
  {Utsumi}}]{Imamura2004}%
  \BibitemOpen
  \bibfield  {author} {\bibinfo {author} {\bibfnamefont {Hiroshi}\ \bibnamefont
  {Imamura}}, \bibinfo {author} {\bibfnamefont {Patrick}\ \bibnamefont
  {Bruno}}, \ and\ \bibinfo {author} {\bibfnamefont {Yasuhiro}\ \bibnamefont
  {Utsumi}},\ }\bibfield  {title} {\enquote {\bibinfo {title} {Twisted exchange
  interaction between localized spins embedded in a one- or two-dimensional
  electron gas with {Rashba} spin-orbit coupling},}\ }\href {\doibase
  10.1103/PhysRevB.69.121303} {\bibfield  {journal} {\bibinfo  {journal} {Phys.
  Rev. B}\ }\textbf {\bibinfo {volume} {69}},\ \bibinfo {pages} {121303(R)}
  (\bibinfo {year} {2004})}\BibitemShut {NoStop}%
\bibitem [{\citenamefont {Kikuchi}\ \emph {et~al.}(2016)\citenamefont
  {Kikuchi}, \citenamefont {Koretsune}, \citenamefont {Arita},\ and\
  \citenamefont {Tatara}}]{Kikuchi2016}%
  \BibitemOpen
  \bibfield  {author} {\bibinfo {author} {\bibfnamefont {Toru}\ \bibnamefont
  {Kikuchi}}, \bibinfo {author} {\bibfnamefont {Takashi}\ \bibnamefont
  {Koretsune}}, \bibinfo {author} {\bibfnamefont {Ryotaro}\ \bibnamefont
  {Arita}}, \ and\ \bibinfo {author} {\bibfnamefont {Gen}\ \bibnamefont
  {Tatara}},\ }\bibfield  {title} {\enquote {\bibinfo {title}
  {{Dzyaloshinskii-Moriya} interaction as a consequence of a {Doppler} shift
  due to spin-orbit-induced intrinsic spin current},}\ }\href {\doibase
  10.1103/PhysRevLett.116.247201} {\bibfield  {journal} {\bibinfo  {journal}
  {Phys. Rev. Lett.}\ }\textbf {\bibinfo {volume} {116}},\ \bibinfo {pages}
  {247201} (\bibinfo {year} {2016})}\BibitemShut {NoStop}%
\bibitem [{\citenamefont {Bode}\ \emph {et~al.}(2007)\citenamefont {Bode},
  \citenamefont {Heide}, \citenamefont {Von~Bergmann}, \citenamefont
  {Ferriani}, \citenamefont {Heinze}, \citenamefont {Bihlmayer}, \citenamefont
  {Kubetzka}, \citenamefont {Pietzsch}, \citenamefont {Bl{\"u}gel},\ and\
  \citenamefont {Wiesendanger}}]{Bode2007}%
  \BibitemOpen
  \bibfield  {author} {\bibinfo {author} {\bibfnamefont {Matthias}\
  \bibnamefont {Bode}}, \bibinfo {author} {\bibfnamefont {M}~\bibnamefont
  {Heide}}, \bibinfo {author} {\bibfnamefont {K}~\bibnamefont {Von~Bergmann}},
  \bibinfo {author} {\bibfnamefont {P}~\bibnamefont {Ferriani}}, \bibinfo
  {author} {\bibfnamefont {S}~\bibnamefont {Heinze}}, \bibinfo {author}
  {\bibfnamefont {G}~\bibnamefont {Bihlmayer}}, \bibinfo {author}
  {\bibfnamefont {A}~\bibnamefont {Kubetzka}}, \bibinfo {author} {\bibfnamefont
  {O}~\bibnamefont {Pietzsch}}, \bibinfo {author} {\bibfnamefont
  {S}~\bibnamefont {Bl{\"u}gel}}, \ and\ \bibinfo {author} {\bibfnamefont
  {R}~\bibnamefont {Wiesendanger}},\ }\bibfield  {title} {\enquote {\bibinfo
  {title} {Chiral magnetic order at surfaces driven by inversion asymmetry},}\
  }\href {\doibase 10.1038/nature05802} {\bibfield  {journal} {\bibinfo
  {journal} {Nature}\ }\textbf {\bibinfo {volume} {447}},\ \bibinfo {pages}
  {190--193} (\bibinfo {year} {2007})}\BibitemShut {NoStop}%
\bibitem [{\citenamefont {Emori}\ \emph {et~al.}(2013)\citenamefont {Emori},
  \citenamefont {Bauer}, \citenamefont {Ahn}, \citenamefont {Martinez},\ and\
  \citenamefont {Beach}}]{Emori2013}%
  \BibitemOpen
  \bibfield  {author} {\bibinfo {author} {\bibfnamefont {Satoru}\ \bibnamefont
  {Emori}}, \bibinfo {author} {\bibfnamefont {Uwe}\ \bibnamefont {Bauer}},
  \bibinfo {author} {\bibfnamefont {Sung-Min}\ \bibnamefont {Ahn}}, \bibinfo
  {author} {\bibfnamefont {Eduardo}\ \bibnamefont {Martinez}}, \ and\ \bibinfo
  {author} {\bibfnamefont {Geoffrey~SD}\ \bibnamefont {Beach}},\ }\bibfield
  {title} {\enquote {\bibinfo {title} {Current-driven dynamics of chiral
  ferromagnetic domain walls},}\ }\href {\doibase 10.1038/nmat3675} {\bibfield
  {journal} {\bibinfo  {journal} {Nature Materials}\ }\textbf {\bibinfo
  {volume} {12}},\ \bibinfo {pages} {611--616} (\bibinfo {year}
  {2013})}\BibitemShut {NoStop}%
\bibitem [{\citenamefont {Nagaosa}\ and\ \citenamefont
  {Tokura}(2013)}]{Nagaosa2013}%
  \BibitemOpen
  \bibfield  {author} {\bibinfo {author} {\bibfnamefont {Naoto}\ \bibnamefont
  {Nagaosa}}\ and\ \bibinfo {author} {\bibfnamefont {Yoshinori}\ \bibnamefont
  {Tokura}},\ }\bibfield  {title} {\enquote {\bibinfo {title} {Topological
  properties and dynamics of magnetic skyrmions},}\ }\href {\doibase
  10.1038/nnano.2013.243} {\bibfield  {journal} {\bibinfo  {journal} {Nat.
  Nanotechnol.}\ }\textbf {\bibinfo {volume} {8}},\ \bibinfo {pages} {899--911}
  (\bibinfo {year} {2013})}\BibitemShut {NoStop}%
\bibitem [{\citenamefont {Fert}\ \emph {et~al.}(2017)\citenamefont {Fert},
  \citenamefont {Reyren},\ and\ \citenamefont {Cros}}]{Fert2017}%
  \BibitemOpen
  \bibfield  {author} {\bibinfo {author} {\bibfnamefont {Albert}\ \bibnamefont
  {Fert}}, \bibinfo {author} {\bibfnamefont {Nicolas}\ \bibnamefont {Reyren}},
  \ and\ \bibinfo {author} {\bibfnamefont {Vincent}\ \bibnamefont {Cros}},\
  }\bibfield  {title} {\enquote {\bibinfo {title} {Magnetic skyrmions: advances
  in physics and potential applications},}\ }\href {\doibase
  10.1038/natrevmats.2017.31} {\bibfield  {journal} {\bibinfo  {journal}
  {Nature Reviews Materials}\ }\textbf {\bibinfo {volume} {2}},\ \bibinfo
  {pages} {201731} (\bibinfo {year} {2017})}\BibitemShut {NoStop}%
\bibitem [{\citenamefont {Liechtenstein}\ \emph {et~al.}(1987)\citenamefont
  {Liechtenstein}, \citenamefont {Katsnelson}, \citenamefont {Antropov},\ and\
  \citenamefont {Gubanov}}]{Liechtenstein1987}%
  \BibitemOpen
  \bibfield  {author} {\bibinfo {author} {\bibfnamefont {A~Il}\ \bibnamefont
  {Liechtenstein}}, \bibinfo {author} {\bibfnamefont {MI}~\bibnamefont
  {Katsnelson}}, \bibinfo {author} {\bibfnamefont {VP}~\bibnamefont
  {Antropov}}, \ and\ \bibinfo {author} {\bibfnamefont {VA}~\bibnamefont
  {Gubanov}},\ }\bibfield  {title} {\enquote {\bibinfo {title} {Local spin
  density functional approach to the theory of exchange interactions in
  ferromagnetic metals and alloys},}\ }\href {\doibase
  10.1016/0304-8853(87)90721-9} {\bibfield  {journal} {\bibinfo  {journal} {J.
  Magn. Magn. Mater.}\ }\textbf {\bibinfo {volume} {67}},\ \bibinfo {pages}
  {65--74} (\bibinfo {year} {1987})}\BibitemShut {NoStop}%
\bibitem [{\citenamefont {Udvardi}\ \emph {et~al.}(2003)\citenamefont
  {Udvardi}, \citenamefont {Szunyogh}, \citenamefont {Palot{\'a}s},\ and\
  \citenamefont {Weinberger}}]{Udvardi2003}%
  \BibitemOpen
  \bibfield  {author} {\bibinfo {author} {\bibfnamefont {Laszlo}\ \bibnamefont
  {Udvardi}}, \bibinfo {author} {\bibfnamefont {Laszlo}\ \bibnamefont
  {Szunyogh}}, \bibinfo {author} {\bibfnamefont {K}~\bibnamefont
  {Palot{\'a}s}}, \ and\ \bibinfo {author} {\bibfnamefont {Peter}\ \bibnamefont
  {Weinberger}},\ }\bibfield  {title} {\enquote {\bibinfo {title}
  {First-principles relativistic study of spin waves in thin magnetic films},}\
  }\href {\doibase 10.1103/PhysRevB.68.104436} {\bibfield  {journal} {\bibinfo
  {journal} {Phys. Rev. B}\ }\textbf {\bibinfo {volume} {68}},\ \bibinfo
  {pages} {104436} (\bibinfo {year} {2003})}\BibitemShut {NoStop}%
\bibitem [{\citenamefont {Ebert}\ and\ \citenamefont
  {Mankovsky}(2009)}]{Ebert2009}%
  \BibitemOpen
  \bibfield  {author} {\bibinfo {author} {\bibfnamefont {H}~\bibnamefont
  {Ebert}}\ and\ \bibinfo {author} {\bibfnamefont {S}~\bibnamefont
  {Mankovsky}},\ }\bibfield  {title} {\enquote {\bibinfo {title} {Anisotropic
  exchange coupling in diluted magnetic semiconductors: Ab initio spin-density
  functional theory},}\ }\href {http://prb.aps.org/abstract/PRB/v79/i4/e045209}
  {\bibfield  {journal} {\bibinfo  {journal} {Phys. Rev. B}\ }\textbf {\bibinfo
  {volume} {79}},\ \bibinfo {pages} {045209} (\bibinfo {year}
  {2009})}\BibitemShut {NoStop}%
\bibitem [{\citenamefont {Lounis}\ and\ \citenamefont
  {Dederichs}(2010)}]{Lounis2010a}%
  \BibitemOpen
  \bibfield  {author} {\bibinfo {author} {\bibfnamefont {Samir}\ \bibnamefont
  {Lounis}}\ and\ \bibinfo {author} {\bibfnamefont {Peter~H.}\ \bibnamefont
  {Dederichs}},\ }\bibfield  {title} {\enquote {\bibinfo {title} {Mapping the
  magnetic exchange interactions from first principles: Anisotropy anomaly and
  application to {Fe}, {Ni}, and {Co}},}\ }\href {\doibase
  10.1103/PhysRevB.82.180404} {\bibfield  {journal} {\bibinfo  {journal} {Phys.
  Rev. B}\ }\textbf {\bibinfo {volume} {82}},\ \bibinfo {pages} {180404(R)}
  (\bibinfo {year} {2010})}\BibitemShut {NoStop}%
\bibitem [{\citenamefont {Szilva}\ \emph {et~al.}(2013)\citenamefont {Szilva},
  \citenamefont {Costa}, \citenamefont {Bergman}, \citenamefont {Szunyogh},
  \citenamefont {Nordstr{\"o}m},\ and\ \citenamefont {Eriksson}}]{Szilva2013}%
  \BibitemOpen
  \bibfield  {author} {\bibinfo {author} {\bibfnamefont {Attila}\ \bibnamefont
  {Szilva}}, \bibinfo {author} {\bibfnamefont {M}~\bibnamefont {Costa}},
  \bibinfo {author} {\bibfnamefont {Anders}\ \bibnamefont {Bergman}}, \bibinfo
  {author} {\bibfnamefont {L}~\bibnamefont {Szunyogh}}, \bibinfo {author}
  {\bibfnamefont {Lars}\ \bibnamefont {Nordstr{\"o}m}}, \ and\ \bibinfo
  {author} {\bibfnamefont {Olle}\ \bibnamefont {Eriksson}},\ }\bibfield
  {title} {\enquote {\bibinfo {title} {Interatomic exchange interactions for
  finite-temperature magnetism and nonequilibrium spin dynamics},}\ }\href
  {\doibase 10.1103/PhysRevLett.111.127204} {\bibfield  {journal} {\bibinfo
  {journal} {Phys Rev Lett}\ }\textbf {\bibinfo {volume} {111}},\ \bibinfo
  {pages} {127204} (\bibinfo {year} {2013})}\BibitemShut {NoStop}%
\bibitem [{\citenamefont {Szilva}\ \emph {et~al.}(2017)\citenamefont {Szilva},
  \citenamefont {Thonig}, \citenamefont {Bessarab}, \citenamefont {Kvashnin},
  \citenamefont {Rodrigues}, \citenamefont {Cardias}, \citenamefont {Pereiro},
  \citenamefont {Nordstr\"om}, \citenamefont {Bergman}, \citenamefont
  {Klautau},\ and\ \citenamefont {Eriksson}}]{Szilva2017}%
  \BibitemOpen
  \bibfield  {author} {\bibinfo {author} {\bibfnamefont {A.}~\bibnamefont
  {Szilva}}, \bibinfo {author} {\bibfnamefont {D.}~\bibnamefont {Thonig}},
  \bibinfo {author} {\bibfnamefont {P.~F.}\ \bibnamefont {Bessarab}}, \bibinfo
  {author} {\bibfnamefont {Y.~O.}\ \bibnamefont {Kvashnin}}, \bibinfo {author}
  {\bibfnamefont {D.~C.~M.}\ \bibnamefont {Rodrigues}}, \bibinfo {author}
  {\bibfnamefont {R.}~\bibnamefont {Cardias}}, \bibinfo {author} {\bibfnamefont
  {M.}~\bibnamefont {Pereiro}}, \bibinfo {author} {\bibfnamefont
  {L.}~\bibnamefont {Nordstr\"om}}, \bibinfo {author} {\bibfnamefont
  {A.}~\bibnamefont {Bergman}}, \bibinfo {author} {\bibfnamefont {A.~B.}\
  \bibnamefont {Klautau}}, \ and\ \bibinfo {author} {\bibfnamefont
  {O.}~\bibnamefont {Eriksson}},\ }\bibfield  {title} {\enquote {\bibinfo
  {title} {Theory of noncollinear interactions beyond {Heisenberg} exchange:
  Applications to bcc {Fe}},}\ }\href {\doibase 10.1103/PhysRevB.96.144413}
  {\bibfield  {journal} {\bibinfo  {journal} {Phys. Rev. B}\ }\textbf {\bibinfo
  {volume} {96}},\ \bibinfo {pages} {144413} (\bibinfo {year}
  {2017})}\BibitemShut {NoStop}%
\bibitem [{\citenamefont {Mryasov}\ \emph {et~al.}(1996)\citenamefont
  {Mryasov}, \citenamefont {Freeman},\ and\ \citenamefont
  {Liechtenstein}}]{Mryasov1996}%
  \BibitemOpen
  \bibfield  {author} {\bibinfo {author} {\bibfnamefont {O.~N.}\ \bibnamefont
  {Mryasov}}, \bibinfo {author} {\bibfnamefont {A.~J.}\ \bibnamefont
  {Freeman}}, \ and\ \bibinfo {author} {\bibfnamefont {A.~I.}\ \bibnamefont
  {Liechtenstein}},\ }\bibfield  {title} {\enquote {\bibinfo {title} {Theory of
  non-{Heisenberg exchange}: Results for localized and itinerant magnets},}\
  }\href {\doibase 10.1063/1.361678} {\bibfield  {journal} {\bibinfo  {journal}
  {J. Appl. Phys.}\ }\textbf {\bibinfo {volume} {79}},\ \bibinfo {pages}
  {4805--4807} (\bibinfo {year} {1996})}\BibitemShut {NoStop}%
\bibitem [{\citenamefont {Mankovsky}\ \emph {et~al.}(2020)\citenamefont
  {Mankovsky}, \citenamefont {Polesya},\ and\ \citenamefont
  {Ebert}}]{Mankovsky2020}%
  \BibitemOpen
  \bibfield  {author} {\bibinfo {author} {\bibfnamefont {S.}~\bibnamefont
  {Mankovsky}}, \bibinfo {author} {\bibfnamefont {S.}~\bibnamefont {Polesya}},
  \ and\ \bibinfo {author} {\bibfnamefont {H.}~\bibnamefont {Ebert}},\
  }\bibfield  {title} {\enquote {\bibinfo {title} {Extension of the standard
  {Heisenberg Hamiltonian} to multispin exchange interactions},}\ }\href
  {\doibase 10.1103/PhysRevB.101.174401} {\bibfield  {journal} {\bibinfo
  {journal} {Phys. Rev. B}\ }\textbf {\bibinfo {volume} {101}},\ \bibinfo
  {pages} {174401} (\bibinfo {year} {2020})}\BibitemShut {NoStop}%
\bibitem [{\citenamefont {Lounis}(2020)}]{Lounis2020}%
  \BibitemOpen
  \bibfield  {author} {\bibinfo {author} {\bibfnamefont {Samir}\ \bibnamefont
  {Lounis}},\ }\bibfield  {title} {\enquote {\bibinfo {title}
  {Multiple-scattering approach for multi-spin chiral magnetic interactions:
  application to the one- and two-dimensional {Rashba} electron gas},}\ }\href
  {\doibase 10.1088/1367-2630/abb514} {\bibfield  {journal} {\bibinfo
  {journal} {New Journal of Physics}\ }\textbf {\bibinfo {volume} {22}},\
  \bibinfo {pages} {103003} (\bibinfo {year} {2020})}\BibitemShut {NoStop}%
\bibitem [{\citenamefont {Drautz}\ and\ \citenamefont
  {F{\"a}hnle}(2004)}]{Drautz2004}%
  \BibitemOpen
  \bibfield  {author} {\bibinfo {author} {\bibfnamefont {R}~\bibnamefont
  {Drautz}}\ and\ \bibinfo {author} {\bibfnamefont {M}~\bibnamefont
  {F{\"a}hnle}},\ }\bibfield  {title} {\enquote {\bibinfo {title} {Spin-cluster
  expansion: parametrization of the general adiabatic magnetic energy surface
  with ab initio accuracy},}\ }\href {\doibase 10.1103/PhysRevB.69.104404}
  {\bibfield  {journal} {\bibinfo  {journal} {Physical Review B}\ }\textbf
  {\bibinfo {volume} {69}},\ \bibinfo {pages} {104404} (\bibinfo {year}
  {2004})}\BibitemShut {NoStop}%
\bibitem [{\citenamefont {Drautz}\ and\ \citenamefont
  {F{\"a}hnle}(2005)}]{Drautz2005}%
  \BibitemOpen
  \bibfield  {author} {\bibinfo {author} {\bibfnamefont {Ralf}\ \bibnamefont
  {Drautz}}\ and\ \bibinfo {author} {\bibfnamefont {Manfred}\ \bibnamefont
  {F{\"a}hnle}},\ }\bibfield  {title} {\enquote {\bibinfo {title}
  {Parametrization of the magnetic energy at the atomic level},}\ }\href
  {\doibase 10.1103/PhysRevB.72.212405} {\bibfield  {journal} {\bibinfo
  {journal} {Physical Review B}\ }\textbf {\bibinfo {volume} {72}},\ \bibinfo
  {pages} {212405} (\bibinfo {year} {2005})}\BibitemShut {NoStop}%
\bibitem [{\citenamefont {Antal}\ \emph {et~al.}(2008)\citenamefont {Antal},
  \citenamefont {Lazarovits}, \citenamefont {Udvardi}, \citenamefont
  {Szunyogh}, \citenamefont {{\'U}jfalussy},\ and\ \citenamefont
  {Weinberger}}]{Antal2008}%
  \BibitemOpen
  \bibfield  {author} {\bibinfo {author} {\bibfnamefont {A}~\bibnamefont
  {Antal}}, \bibinfo {author} {\bibfnamefont {B}~\bibnamefont {Lazarovits}},
  \bibinfo {author} {\bibfnamefont {L}~\bibnamefont {Udvardi}}, \bibinfo
  {author} {\bibfnamefont {L}~\bibnamefont {Szunyogh}}, \bibinfo {author}
  {\bibfnamefont {B}~\bibnamefont {{\'U}jfalussy}}, \ and\ \bibinfo {author}
  {\bibfnamefont {P}~\bibnamefont {Weinberger}},\ }\bibfield  {title} {\enquote
  {\bibinfo {title} {First-principles calculations of spin interactions and the
  magnetic ground states of {Cr} trimers on {Au}(111)},}\ }\href {\doibase
  10.1103/PhysRevB.77.174429} {\bibfield  {journal} {\bibinfo  {journal} {Phys.
  Rev. B}\ }\textbf {\bibinfo {volume} {77}},\ \bibinfo {pages} {174429}
  (\bibinfo {year} {2008})}\BibitemShut {NoStop}%
\bibitem [{\citenamefont {Singer}\ \emph {et~al.}(2011)\citenamefont {Singer},
  \citenamefont {Dietermann},\ and\ \citenamefont {F\"ahnle}}]{Singer2011}%
  \BibitemOpen
  \bibfield  {author} {\bibinfo {author} {\bibfnamefont {R.}~\bibnamefont
  {Singer}}, \bibinfo {author} {\bibfnamefont {F.}~\bibnamefont {Dietermann}},
  \ and\ \bibinfo {author} {\bibfnamefont {M.}~\bibnamefont {F\"ahnle}},\
  }\bibfield  {title} {\enquote {\bibinfo {title} {Spin interactions in bcc and
  fcc {Fe} beyond the {Heisenberg} model},}\ }\href {\doibase
  10.1103/PhysRevLett.107.017204} {\bibfield  {journal} {\bibinfo  {journal}
  {Phys. Rev. Lett.}\ }\textbf {\bibinfo {volume} {107}},\ \bibinfo {pages}
  {017204} (\bibinfo {year} {2011})}\BibitemShut {NoStop}%
\bibitem [{\citenamefont {Brinker}\ \emph {et~al.}(2019)\citenamefont
  {Brinker}, \citenamefont {{dos Santos Dias}},\ and\ \citenamefont
  {Lounis}}]{Brinker2019}%
  \BibitemOpen
  \bibfield  {author} {\bibinfo {author} {\bibfnamefont {Sascha}\ \bibnamefont
  {Brinker}}, \bibinfo {author} {\bibfnamefont {Manuel}\ \bibnamefont {{dos
  Santos Dias}}}, \ and\ \bibinfo {author} {\bibfnamefont {Samir}\ \bibnamefont
  {Lounis}},\ }\bibfield  {title} {\enquote {\bibinfo {title} {The chiral
  biquadratic pair interaction},}\ }\href {\doibase 10.1088/1367-2630/ab35c9}
  {\bibfield  {journal} {\bibinfo  {journal} {New J. Phys.}\ }\textbf {\bibinfo
  {volume} {21}},\ \bibinfo {pages} {083015} (\bibinfo {year}
  {2019})}\BibitemShut {NoStop}%
\bibitem [{\citenamefont {Brinker}\ \emph {et~al.}(2020)\citenamefont
  {Brinker}, \citenamefont {dos Santos~Dias},\ and\ \citenamefont
  {Lounis}}]{Brinker2020a}%
  \BibitemOpen
  \bibfield  {author} {\bibinfo {author} {\bibfnamefont {Sascha}\ \bibnamefont
  {Brinker}}, \bibinfo {author} {\bibfnamefont {Manuel}\ \bibnamefont {dos
  Santos~Dias}}, \ and\ \bibinfo {author} {\bibfnamefont {Samir}\ \bibnamefont
  {Lounis}},\ }\bibfield  {title} {\enquote {\bibinfo {title} {Prospecting
  chiral multisite interactions in prototypical magnetic systems},}\ }\href
  {\doibase 10.1103/PhysRevResearch.2.033240} {\bibfield  {journal} {\bibinfo
  {journal} {Phys. Rev. Research}\ }\textbf {\bibinfo {volume} {2}},\ \bibinfo
  {pages} {033240} (\bibinfo {year} {2020})}\BibitemShut {NoStop}%
\bibitem [{\citenamefont {L\'aszl\'offy}\ \emph {et~al.}(2019)\citenamefont
  {L\'aszl\'offy}, \citenamefont {R\'ozsa}, \citenamefont {Palot\'as},
  \citenamefont {Udvardi},\ and\ \citenamefont {Szunyogh}}]{Laszloffy2019}%
  \BibitemOpen
  \bibfield  {author} {\bibinfo {author} {\bibfnamefont {A.}~\bibnamefont
  {L\'aszl\'offy}}, \bibinfo {author} {\bibfnamefont {L.}~\bibnamefont
  {R\'ozsa}}, \bibinfo {author} {\bibfnamefont {K.}~\bibnamefont {Palot\'as}},
  \bibinfo {author} {\bibfnamefont {L.}~\bibnamefont {Udvardi}}, \ and\
  \bibinfo {author} {\bibfnamefont {L.}~\bibnamefont {Szunyogh}},\ }\bibfield
  {title} {\enquote {\bibinfo {title} {Magnetic structure of monatomic {Fe}
  chains on {Re}(0001): Emergence of chiral multispin interactions},}\ }\href
  {\doibase 10.1103/PhysRevB.99.184430} {\bibfield  {journal} {\bibinfo
  {journal} {Phys. Rev. B}\ }\textbf {\bibinfo {volume} {99}},\ \bibinfo
  {pages} {184430} (\bibinfo {year} {2019})}\BibitemShut {NoStop}%
\bibitem [{\citenamefont {Grytsiuk}\ \emph {et~al.}(2020)\citenamefont
  {Grytsiuk}, \citenamefont {Hanke}, \citenamefont {Hoffmann}, \citenamefont
  {Bouaziz}, \citenamefont {Gomonay}, \citenamefont {Bihlmayer}, \citenamefont
  {Lounis}, \citenamefont {Mokrousov},\ and\ \citenamefont
  {Bl\"ugel}}]{Grytsiuk2020}%
  \BibitemOpen
  \bibfield  {author} {\bibinfo {author} {\bibfnamefont {S.}~\bibnamefont
  {Grytsiuk}}, \bibinfo {author} {\bibfnamefont {J.-P.}\ \bibnamefont {Hanke}},
  \bibinfo {author} {\bibfnamefont {M.}~\bibnamefont {Hoffmann}}, \bibinfo
  {author} {\bibfnamefont {J.}~\bibnamefont {Bouaziz}}, \bibinfo {author}
  {\bibfnamefont {O.}~\bibnamefont {Gomonay}}, \bibinfo {author} {\bibfnamefont
  {G.}~\bibnamefont {Bihlmayer}}, \bibinfo {author} {\bibfnamefont
  {S.}~\bibnamefont {Lounis}}, \bibinfo {author} {\bibfnamefont
  {Y.}~\bibnamefont {Mokrousov}}, \ and\ \bibinfo {author} {\bibfnamefont
  {S.}~\bibnamefont {Bl\"ugel}},\ }\bibfield  {title} {\enquote {\bibinfo
  {title} {Topological-chiral magnetic interactions driven by emergent orbital
  magnetism},}\ }\href {\doibase 10.1038/s41467-019-14030-3} {\bibfield
  {journal} {\bibinfo  {journal} {Nat. Commun.}\ }\textbf {\bibinfo {volume}
  {11}},\ \bibinfo {pages} {511} (\bibinfo {year} {2020})}\BibitemShut
  {NoStop}%
\bibitem [{\citenamefont {{Cardias}}\ \emph {et~al.}(2020)\citenamefont
  {{Cardias}}, \citenamefont {{Bergman}}, \citenamefont {{Szilva}},
  \citenamefont {{Kvashnin}}, \citenamefont {{Fransson}}, \citenamefont
  {{Klautau}}, \citenamefont {{Eriksson}},\ and\ \citenamefont
  {{Nordstr{\"o}m}}}]{Cardias2020}%
  \BibitemOpen
  \bibfield  {author} {\bibinfo {author} {\bibfnamefont {Ramon}\ \bibnamefont
  {{Cardias}}}, \bibinfo {author} {\bibfnamefont {Anders}\ \bibnamefont
  {{Bergman}}}, \bibinfo {author} {\bibfnamefont {Attila}\ \bibnamefont
  {{Szilva}}}, \bibinfo {author} {\bibfnamefont {Yaroslav~O.}\ \bibnamefont
  {{Kvashnin}}}, \bibinfo {author} {\bibfnamefont {Jonas}\ \bibnamefont
  {{Fransson}}}, \bibinfo {author} {\bibfnamefont {Angela~B.}\ \bibnamefont
  {{Klautau}}}, \bibinfo {author} {\bibfnamefont {Olle}\ \bibnamefont
  {{Eriksson}}}, \ and\ \bibinfo {author} {\bibfnamefont {Lars}\ \bibnamefont
  {{Nordstr{\"o}m}}},\ }\href@noop {} {\enquote {\bibinfo {title}
  {{Dzyaloshinskii-Moriya interaction in absence of spin-orbit coupling}},}\ }
  (\bibinfo {year} {2020}),\ \Eprint {http://arxiv.org/abs/2003.04680}
  {arXiv:2003.04680 [cond-mat.mtrl-sci]} \BibitemShut {NoStop}%
\bibitem [{\citenamefont {Cardias}\ \emph {et~al.}(2020)\citenamefont
  {Cardias}, \citenamefont {Szilva}, \citenamefont {Bezerra-Neto},
  \citenamefont {Ribeiro}, \citenamefont {Bergman}, \citenamefont {Kvashnin},
  \citenamefont {Fransson}, \citenamefont {Klautau}, \citenamefont {Eriksson},\
  and\ \citenamefont {Nordstr\"om}}]{Cardias2020a}%
  \BibitemOpen
  \bibfield  {author} {\bibinfo {author} {\bibfnamefont {R.}~\bibnamefont
  {Cardias}}, \bibinfo {author} {\bibfnamefont {A.}~\bibnamefont {Szilva}},
  \bibinfo {author} {\bibfnamefont {M.~M.}\ \bibnamefont {Bezerra-Neto}},
  \bibinfo {author} {\bibfnamefont {M.~S.}\ \bibnamefont {Ribeiro}}, \bibinfo
  {author} {\bibfnamefont {A.}~\bibnamefont {Bergman}}, \bibinfo {author}
  {\bibfnamefont {Y.~O.}\ \bibnamefont {Kvashnin}}, \bibinfo {author}
  {\bibfnamefont {J.}~\bibnamefont {Fransson}}, \bibinfo {author}
  {\bibfnamefont {A.~B.}\ \bibnamefont {Klautau}}, \bibinfo {author}
  {\bibfnamefont {O.}~\bibnamefont {Eriksson}}, \ and\ \bibinfo {author}
  {\bibfnamefont {L.}~\bibnamefont {Nordstr\"om}},\ }\bibfield  {title}
  {\enquote {\bibinfo {title} {First-principles {Dzyaloshinskii-Moriya}
  interaction in a non-collinear framework},}\ }\href {\doibase
  10.1038/s41598-020-77219-3} {\bibfield  {journal} {\bibinfo  {journal} {Sci.
  Rep.}\ }\textbf {\bibinfo {volume} {10}},\ \bibinfo {pages} {20339} (\bibinfo
  {year} {2020})}\BibitemShut {NoStop}%
\bibitem [{\citenamefont {Papanikolaou}\ \emph {et~al.}(2002)\citenamefont
  {Papanikolaou}, \citenamefont {Zeller},\ and\ \citenamefont
  {Dederichs}}]{Papanikolaou:2002}%
  \BibitemOpen
  \bibfield  {author} {\bibinfo {author} {\bibfnamefont {N.}~\bibnamefont
  {Papanikolaou}}, \bibinfo {author} {\bibfnamefont {R.}~\bibnamefont
  {Zeller}}, \ and\ \bibinfo {author} {\bibfnamefont {P.~H.}\ \bibnamefont
  {Dederichs}},\ }\bibfield  {title} {\enquote {\bibinfo {title} {Conceptual
  improvements of the {KKR} method},}\ }\href {\doibase
  10.1088/0953-8984/14/11/304} {\bibfield  {journal} {\bibinfo  {journal}
  {Journal of Physics: Condensed Matter}\ }\textbf {\bibinfo {volume} {14}},\
  \bibinfo {pages} {2799--2823} (\bibinfo {year} {2002})}\BibitemShut {NoStop}%
\bibitem [{\citenamefont {Bauer}(2014)}]{Bauer2014}%
  \BibitemOpen
  \bibfield  {author} {\bibinfo {author} {\bibfnamefont {D.~S.~G.}\
  \bibnamefont {Bauer}},\ }\emph {\bibinfo {title} {Development of a
  relativistic full-potential first-principles multiple scattering Green
  function method applied to complex magnetic textures of nano structures at
  surfaces}},\ \href@noop {} {Ph.D. thesis},\ \bibinfo  {school} {RWTH Aachen}
  (\bibinfo {year} {2014})\BibitemShut {NoStop}%
\bibitem [{\citenamefont {Vosko}\ \emph {et~al.}(1980)\citenamefont {Vosko},
  \citenamefont {Wilk},\ and\ \citenamefont {Nusair}}]{Vosko1980}%
  \BibitemOpen
  \bibfield  {author} {\bibinfo {author} {\bibfnamefont {SH}~\bibnamefont
  {Vosko}}, \bibinfo {author} {\bibfnamefont {L}~\bibnamefont {Wilk}}, \ and\
  \bibinfo {author} {\bibfnamefont {M}~\bibnamefont {Nusair}},\ }\bibfield
  {title} {\enquote {\bibinfo {title} {Accurate spin-dependent electron liquid
  correlation energies for local spin density calculations: a critical
  analysis},}\ }\href {\doibase 10.1139/p80-159} {\bibfield  {journal}
  {\bibinfo  {journal} {Can. J. Phys.}\ }\textbf {\bibinfo {volume} {58}},\
  \bibinfo {pages} {1200--1211} (\bibinfo {year} {1980})}\BibitemShut {NoStop}%
\bibitem [{\citenamefont {Tomiyoshi}(1982)}]{Tomiyoshi1982}%
  \BibitemOpen
  \bibfield  {author} {\bibinfo {author} {\bibfnamefont {Sh\={o}ichi}\
  \bibnamefont {Tomiyoshi}},\ }\bibfield  {title} {\enquote {\bibinfo {title}
  {Polarized neutron diffraction study of the spin structure of {Mn$_3$Sn}},}\
  }\href {\doibase 10.1143/JPSJ.51.803} {\bibfield  {journal} {\bibinfo
  {journal} {Journal of the Physical Society of Japan}\ }\textbf {\bibinfo
  {volume} {51}},\ \bibinfo {pages} {803--810} (\bibinfo {year}
  {1982})}\BibitemShut {NoStop}%
\bibitem [{\citenamefont {Ny\'ari}\ \emph {et~al.}(2019)\citenamefont
  {Ny\'ari}, \citenamefont {De\'ak},\ and\ \citenamefont
  {Szunyogh}}]{Nyari2019}%
  \BibitemOpen
  \bibfield  {author} {\bibinfo {author} {\bibfnamefont {B.}~\bibnamefont
  {Ny\'ari}}, \bibinfo {author} {\bibfnamefont {A.}~\bibnamefont {De\'ak}}, \
  and\ \bibinfo {author} {\bibfnamefont {L.}~\bibnamefont {Szunyogh}},\
  }\bibfield  {title} {\enquote {\bibinfo {title} {Weak ferromagnetism in
  hexagonal {$\mathrm{Mn}_3Z$} alloys
  {$(Z=\mathrm{Sn},\mathrm{Ge},\mathrm{Ga})$}},}\ }\href {\doibase
  10.1103/PhysRevB.100.144412} {\bibfield  {journal} {\bibinfo  {journal}
  {Phys. Rev. B}\ }\textbf {\bibinfo {volume} {100}},\ \bibinfo {pages}
  {144412} (\bibinfo {year} {2019})}\BibitemShut {NoStop}%
\bibitem [{\citenamefont {Hoffmann}\ and\ \citenamefont
  {Bl\"ugel}(2020)}]{Hoffmann2020}%
  \BibitemOpen
  \bibfield  {author} {\bibinfo {author} {\bibfnamefont {Markus}\ \bibnamefont
  {Hoffmann}}\ and\ \bibinfo {author} {\bibfnamefont {Stefan}\ \bibnamefont
  {Bl\"ugel}},\ }\bibfield  {title} {\enquote {\bibinfo {title} {Systematic
  derivation of realistic spin models for beyond-{Heisenberg} solids},}\ }\href
  {\doibase 10.1103/PhysRevB.101.024418} {\bibfield  {journal} {\bibinfo
  {journal} {Phys. Rev. B}\ }\textbf {\bibinfo {volume} {101}},\ \bibinfo
  {pages} {024418} (\bibinfo {year} {2020})}\BibitemShut {NoStop}%
\bibitem [{\citenamefont {{J\"{u}lich Supercomputing Centre}}(2018)}]{jureca}%
  \BibitemOpen
  \bibfield  {author} {\bibinfo {author} {\bibnamefont {{J\"{u}lich
  Supercomputing Centre}}},\ }\bibfield  {title} {\enquote {\bibinfo {title}
  {{JURECA: Modular supercomputer at J\"{u}lich Supercomputing Centre}},}\
  }\href {\doibase 10.17815/jlsrf-4-121-1} {\bibfield  {journal} {\bibinfo
  {journal} {Journal of large-scale research facilities}\ }\textbf {\bibinfo
  {volume} {4}} (\bibinfo {year} {2018}),\ 10.17815/jlsrf-4-121-1}\BibitemShut
  {NoStop}%
\end{thebibliography}%

\end{document}